\documentclass[conference]{IEEEtran}
\IEEEoverridecommandlockouts
\usepackage[top=0.75in,bottom=1.05in,left=0.75in,right=0.75in]{geometry}

\usepackage{cite}
\usepackage{caption}
\usepackage{amsmath,amssymb,amsfonts}
\usepackage{algorithmic}
\usepackage{graphicx}
\usepackage{textcomp}
\usepackage{xcolor}
\usepackage{cleveref}
\usepackage{booktabs}
\usepackage{subcaption}

\def\BibTeX{{\rm B\kern-.05em{\sc i\kern-.025em b}\kern-.08em
    T\kern-.1667em\lower.7ex\hbox{E}\kern-.125emX}}
\begin{document}

\title{Joint Scheduling and Resource Allocation in mmWave IAB Networks Using Deep RL}

\author{\IEEEauthorblockN{Maryam Abbasalizadeh}
\IEEEauthorblockA{
\textit{University of Massachusetts Lowell}\\
Lowell, MA, USA \\
maryam\_abbasalizadeh@student.uml.edu}
\and
\IEEEauthorblockN{Sashank Narain}
\IEEEauthorblockA{
\textit{University of Massachusetts Lowell}\\
Lowell, MA, USA \\
sashank\_narain@uml.edu}
}

\maketitle
\begin{abstract}

Integrated Access and Backhaul (IAB) is critical for dense 5G and beyond deployments, especially in mmWave bands where fiber backhaul is infeasible. We propose a novel Deep Reinforcement Learning (DRL) framework for joint link scheduling and resource slicing in dynamic, interference-prone IAB networks. Our method integrates a greedy Double Deep Q-Network (DDQN) scheduler to activate access and backhaul links based on traffic and topology, with a multi-agent DDQN allocator for bandwidth and antenna assignment across network slices. This decentralized approach respects strict antenna constraints and supports concurrent scheduling across heterogeneous links. Evaluations across 96 dynamic topologies show 99.84\% scheduling accuracy and 20.90\% throughput improvement over baselines. The framework’s efficient operation and adaptability make it suitable for dynamic and resource-constrained deployments, where fast link scheduling and autonomous backhaul coordination are vital.

\end{abstract}

%The framework’s low-latency operation and adaptability make it suitable for tactical and time-sensitive deployments, where autonomous backhaul coordination is vital.


\section{Introduction}

The rapid growth in mobile data demand, driven by augmented reality, autonomous vehicles, and massive IoT, is accelerating the rollout of ultra-dense 5G and beyond networks. A central challenge in these deployments is backhaul provisioning, which becomes especially costly and complex in the mmWave spectrum, where high-capacity access is possible but fiber installation is often impractical~\cite{sadovaya2024impact}. To address this, 3GPP introduced Integrated Access and Backhaul (IAB), allowing wireless relay nodes to function as both access and backhaul points by reusing radio resources~\cite{ghodhbane2024delay}. This design enables flexible, cost-effective deployments and reduces reliance on fiber, especially where infrastructure is sparse or rapidly deployable systems are required, such as tactical or emergency-response networks.

However, IAB poses significant challenges, especially in dynamic in-band mmWave environments marked by rapid channel variability, directional transmissions, and limited spectrum and hardware resources~\cite{song2017relay}. Managing interference, scheduling links, and allocating shared resources like bandwidth and antennas across multiple slices is inherently complex~\cite{cho2025resource,ganesh2024impact,morgado2023deep}. Consequently, traditional static or heuristic-based methods often fall short, struggling to adapt to fast-changing traffic demands and topologies, leading to suboptimal performance. Moreover, RF-based backhaul systems below 6 GHz face limitations such as restricted bandwidth and high interference, which worsen in dense IAB settings. While power control mechanisms offer partial relief, their effectiveness is limited by coordination overhead, latency, and unresolved contention issues~\cite{shashidhara2024hierarchical,vu2016joint}.

To address these limitations, we introduce a novel Deep Reinforcement Learning (DRL) framework for joint scheduling and resource slicing in mmWave-enabled IAB networks. The framework integrates a greedy Double Deep Q-Network (DDQN) scheduler, which selects links based on dynamic traffic demand and network topology, with a multi-agent DDQN system that dynamically allocates bandwidth and antenna resources across network slices to satisfy QoS requirements. This architecture enables fine-grained, adaptive control under strict resource constraints and supports concurrent scheduling of UE-to-IAB, IAB-to-IAB, and donor gNB (DgNB)-IAB links. The decentralized framework is well-suited for real-time operation in large, heterogeneous networks. Preliminary evaluations across dynamic network topologies demonstrate promising performance, achieving near-optimal scheduling accuracy (99.84\%) and notable throughput gains (20.90\%) over state-of-the-art baselines. These results highlight the potential of our approach to effectively manage the complexity of joint scheduling and slicing in next-generation wireless backhaul networks.
% \textcolor {blue} {In summary, this paper makes the following contributions.}

% \begin{itemize}

% \item \textcolor {blue} {We propose a novel DRL framework combining greedy and multi-agent DDQN for joint scheduling and resource slicing in dynamic mmWave IAB networks.}

% \item \textcolor {blue} {Extensive simulations demonstrate the effectiveness of our framework, achieving 99.84\% scheduling accuracy and 20.90\% throughput improvement over baselines.
% }

% \end{itemize}

The paper is organized as follows. Section~\ref{sec:relatedwork} reviews related work on link scheduling and resource allocation in 5G networks. Section~\ref{sec:systemdesign} outlines the system model, including network topology and resource abstraction. Section~\ref{sec:systemdesign2} details the proposed DRL-based scheduling and allocation framework. Section~\ref{sec:evaluationresults} presents experimental results under dynamic conditions, and Section~\ref{sec:conclusion} concludes the paper.

\section{Related Work}\label{sec:relatedwork}

\vspace{0.25em}
\noindent\textit{Link Scheduling in 5G and IAB Networks.}
Scheduling in 5G and mmWave-based IAB networks is challenging due to directional links, limited hardware, and variable traffic demands. Song et al.\cite{song2017relay} propose a cooperative multi-hop free-space optical (FSO) backhaul system using CSI and a back-pressure relay selection algorithm to enhance reliability and scheduling. Shashidhara et al.\cite{shashidhara2024hierarchical} present a hierarchical framework combining centralized policy design with distributed execution, yielding high gains in latency and throughput for multi-antenna systems. Qian et al.~\cite{qian2019joint} introduce a virtual link abstraction to model RF chain contention in full-duplex MIMO and develop a CSMA-like distributed scheduler for scalable MAC-layer coordination.

In IAB-specific contexts, Vu et al.\cite{vu2016joint} jointly optimize scheduling, beamforming, and power allocation in full-duplex heterogeneous networks using Lyapunov-based control. Ghodhbane et al.\cite{ghodhbane2024delay} propose a semi-centralized resource management scheme aligned with 3GPP standards, showing latency and throughput improvements in shared-spectrum IAB deployments. Cho et al.\cite{cho2025resource} improve fairness and capacity in dense settings via a two-phase scheduler combining MU-MIMO and NOMA. Sadovaya et al.\cite{sadovaya2024impact} present a mixed-integer nonlinear programming (MINLP) approach for joint resource optimization, factoring in user mobility and dynamic traffic.

\vspace{0.25em}
\noindent\textit{Learning-Based Resource Allocation and Slicing.}
Recent work has applied Deep Reinforcement Learning (DRL) to resource allocation and network slicing in IAB networks. Morgado et al.~\cite{morgado2023deep, morgado2024intelligent} use Double Deep Q-Networks (DDQN) for adaptive backhaul selection and slice-aware routing in dynamic 5G and satellite-integrated systems. Though effective under non-stationary conditions, their approach focuses only on the backhaul layer, neglecting access-side scheduling and antenna constraints. Fabian et al.\cite{fabian2021performance} assess IAB’s performance using backhaul adaptation protocol (BAP)-managed trees but limit their analysis to network-layer evaluations without adaptive control. Complementary work by Ganesh et al.~\cite{ganesh2024impact} highlights the impact of antenna beam patterns on interference and signal integrity, underscoring the need for flexible, direction-aware scheduling in mmWave settings.

\vspace{0.25em}
\noindent\textit{Positioning and Contributions.}
Departing from approaches that isolate scheduling, slicing, or backhaul selection, our work introduces a unified DRL framework for joint scheduling and resource slicing in mmWave-enabled IAB networks. Combining a greedy DDQN scheduler with a multi-agent DDQN allocator enables real-time, fine-grained control of link activation, bandwidth distribution, and antenna assignment across access and backhaul. Our model captures realistic, dynamic topologies over a full-day deployment using 15-minute intervals, reflecting mobility, congestion, and QoS variability across slices. To our knowledge, this is the first work to integrate access scheduling, backhaul coordination, and slice-aware resource provisioning under antenna and bandwidth constraints using DRL in in-band IAB networks.

\section{Background}\label{sec:systemdesign}

\subsection{Network Model}\label{sec:network-structure}

We consider a mmWave-based IAB network comprising multiple IAB nodes wirelessly connected with one another and a donor node $n_0$. This donor node connects to the core network via a wired backhaul. Each node $n_i \in \mathcal{N} = \{n_0, n_1, \ldots, n_{\hat{m}}\}$ is equipped with a set of directional antennas $\mathcal{A}_{n_i} = \{a_{i,1}, \ldots, a_{i,k_i}\}$, where each antenna can support at most one active link per timeslot. The network accommodates three types of links: UE-to-IAB (access), IAB-to-IAB, and IAB-to-donor. All wireless links are assumed to be line-of-sight (LoS), a realistic condition for IAB deployments where nodes are placed at fixed, elevated locations and links are planned to maintain unobstructed backhaul paths.

Communication is conducted in a half-duplex, in-band mode, meaning that access and backhaul transmissions share the same mmWave spectrum. This design choice significantly reduces deployment complexity and cost while improving spectral efficiency. However, it also intensifies interference and leads to heightened competition for limited resources, particularly at the antenna level, posing major challenges for real-time network coordination. These constraints underscore the need for intelligent, scalable scheduling strategies to jointly manage link activation, spectrum sharing, and antenna assignment.

\subsection{Interference-Aware Scheduling Model}\label{sec:link-scheduling}

While highly directional transmissions mitigate interference in mmWave IAB networks, it remains a significant challenge due to side lobe leakage, beam misalignment, and the limited antennas that can be concurrently active at each node. These issues are particularly pronounced in dense deployments, where narrow beams from multiple links may still spatially overlap, increasing the risk of mutual interference. In our model, each node is constrained to activating limited links per time slot, introducing complex trade-offs among link utility, interference potential, and hardware limitations.

Traditional physical-layer techniques, such as those based on channel state information (CSI), signal-to-interference-plus-noise ratio (SINR) thresholds, or adaptive beamforming, can be effective but are typically resource-intensive and poorly suited to real-time operation in dynamic topologies. To enable scalable and responsive scheduling, we employ a high-level abstraction in which each candidate link is assigned a scalar weight that quantifies its scheduling priority. These weights can reflect urgency in QoS and implicitly capture dynamic network conditions, including bandwidth requirements, traffic load, and signal quality degradation.

\subsection{Slice-Aware Resource Allocation Model}\label{sec:resource-allocation}

Following link scheduling, each IAB node must allocate its limited bandwidth and antenna resources to active network slices while meeting their specific QoS requirements. The network supports three slice types—enhanced Mobile Broadband (eMBB), ultra-Reliable Low Latency Communications (uRLLC), and enhanced Machine-Type Communications (eMTC)—each with distinct resource demands.

The network topology and traffic profile are updated every 15 minutes over 24 hours to capture temporal variation in resource usage. At each time step, the system monitors the load on active links and the residual bandwidth and antenna capacity at each base station. When a local base station becomes congested, traffic from one or more slices can be offloaded via wireless backhaul to neighboring nodes, provided sufficient resources are available.

Slice-level demands are expressed as $\text{Slice\_profile}(t, i, \text{sid}, \text{Band}, \text{Antenna}),$ where $t$ is the time index, $i$ is the base station identifier, and \texttt{sid} denotes the slice. The \texttt{Band} and \texttt{Antenna} fields specify the required resources. Residual resources at each base station after link scheduling are defined as $\text{Load\_profile}(t, i, \text{Band}, \text{Antenna}),$ representing the available bandwidth and antenna capacity. These profiles inform the DRL-based resource allocator, enabling dynamic, slice-aware assignment of bandwidth and antennas under realistic IAB constraints.

\section{Methodology}\label{sec:systemdesign2}

\subsection{Deployment Scenario and Slice Configuration}\label{sec:network-topology}

\begin{figure}[t]
\centering
\includegraphics[width=.7\linewidth]{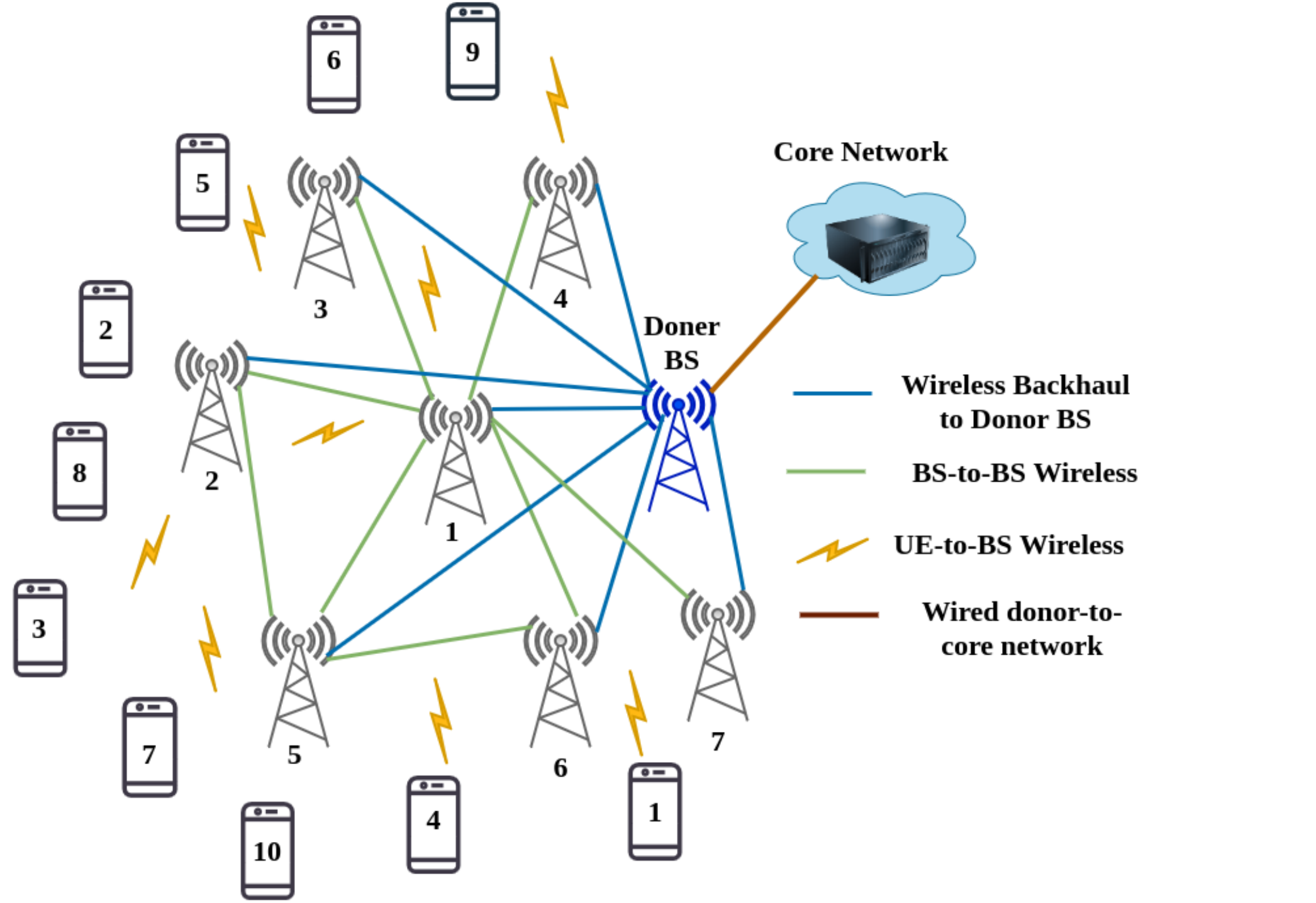}
\caption{Simulated network deployment scenario.}
\label{fig:network_topology}
\end{figure}

Our simulated deployment includes seven base stations (BS1–BS7), ten user equipment (UE1–UE10), and a donor gNB (DgNB) wire-connected to the core network. The donor BS provides wireless backhaul to the others. Each BS has $K = 14$ directional antennas, each supporting one active link per scheduling interval. UEs may associate with any BS, as shown in~\Cref{fig:network_topology}. BS1, designated as a persistently congested node and main aggregation point, maintains links with all other BSs throughout dynamic topology updates, which occur every 15 minutes over 24 hours. The system schedules all active UE-to-BS and BS-to-BS links at each interval to optimize bandwidth and antenna use.

This topology reflects a realistic urban deployment aligned with 3GPP IAB standards, where BSs are wirelessly backhauled to a donor DgNB. The seven BSs introduce inter-BS coordination and interference challenges, while ten UEs generate heterogeneous, time-varying traffic. A persistently overloaded BS (BS1) simulates load hotspots typical of dense urban areas. This setup enables detailed evaluation of slice-aware resource allocation, congestion-sensitive backhaul, and DRL-based adaptation under dynamic conditions.

\subsection{DRL for Link Scheduling}\label{sec:RL-scheduling}

We apply a Double Deep Q-Network (DDQN) framework to perform adaptive link scheduling in dynamic, resource-constrained mmWave IAB networks. The objective is to select an optimal subset of links to activate at each scheduling interval, subject to per-node antenna constraints. Each candidate link $\ell_i$ is characterized by a normalized weight $w_i \in [0,1]$, indicating bandwidth demand or scheduling urgency, and a type indicator $\tau_i \in \{0,1\}$, where $\tau_i = 0$ denotes infrastructure links (BS-to-BS or BS-to-donor), and $\tau_i = 1$ denotes access links (BS-to-UE). Prioritizing infrastructure links supports overall network connectivity, making link type a meaningful secondary ranking criterion.

\vspace{0.2em}
\noindent\textit{Observation and Action Representation:} 
At each time step, the agent observes a compact matrix $S \in \mathbb{R}^{N \times 2}$, where each row encodes a link $\ell_i$ as the pair $(w_i, \tau_i)$. This abstraction captures link demand and role without requiring CSI or explicit topology graphs. The matrix is flattened into a state vector $x \in \mathbb{R}^{2N}$, which serves as input to the Q-network.

The agent's action is to select a feasible subset of links for activation, ensuring no node $n_j \in \mathcal{N}$ exceeds its hardware constraint of $K$ active directional antennas:
$$
\sum_{\ell_i \in \mathcal{A} : \text{src}(\ell_i) = n_j} 1 \leq K, \quad \forall n_j \in \mathcal{N}
$$

Given that the number of valid subsets grows exponentially with $N$, exhaustive search is computationally infeasible. To address this, we adopt a greedy Q-value-guided selection strategy. The agent ranks links by their estimated utility (Q-values) and iteratively selects the highest-ranked feasible links, balancing near-optimality with efficiency.

\vspace{0.2em}
\noindent\textit{Greedy-DDQN Scheduling Policy:}
Candidate links are prioritized by the DRL agent using a two-level ranking rule:
$
\text{Rank}(\ell_i) = (-w_i, \tau_i).
$
Links with higher weights (larger $w_i$) are preferred, and the negative sign ensures prioritization via \texttt{argmin}-based sorting. When weights are equal, infrastructure links ($\tau_i = 0$) are favored for maintaining end-to-end connectivity because failure to schedule an infrastructure link may disrupt downstream access paths for multiple UEs. The agent activates links in ascending rank order, subject to per-node antenna constraints.

\vspace{0.2em}
\noindent\textit{Q-Value Estimation and Training:}
The DDQN architecture estimates the action-value function $Q(s, a; \theta)$, where $s \in \mathbb{R}^{2N}$ is the input state, $a$ an action corresponding to selecting a link, and $\theta$ denotes the network parameters. Each Q-value represents the long-term expected reward of activating link as action $a$ in state $s$, accounting for both immediate demand and future scheduling opportunities. Our system's Q-network consists of two hidden layers with 32 ReLU-activated units each, selected based on experimentation with multiple architectures.

The training process focuses on minimizing the squared Bellman error:
$$
\mathcal{L}(\theta) = \left( y - Q(s, a; \theta) \right)^2
$$
using mini-batch stochastic gradient descent with samples drawn from an experience replay buffer. The target network is also periodically updated by syncing its weights from the online network. To balance exploration and exploitation, an $\epsilon$-greedy policy is adopted during training. The exploration rate $\epsilon_t$ decays over time as:
$$
\epsilon_t = \max\left(\epsilon_{\min}, \epsilon_0 \cdot (\epsilon_{\text{decay}})^t \right)
$$
where $\epsilon_0$ is the initial exploration rate, $\epsilon_{\text{decay}} \in [0, 1]$ controls the decay rate, $\epsilon_{\min}$ ensures persistent exploration, and \( t \) denotes the current training step.

To reduce overestimation bias and improve stability, we maintain two networks: an online network $Q(s, a; \theta)$ for action selection, and a target network $Q(s, a; \theta^-)$ for evaluation. The training target is computed as:
$$
y = \begin{cases}
r, \text{ if } s' \text{ is terminal} \\
r + \gamma Q(s', \arg\max_{a'} Q(s', a'; \theta); \theta^-), \text{ otherwise}
\end{cases}
$$
where $r$ is the immediate reward, $s'$ is the next state, and $\gamma \in [0,1]$ is the discount factor. The online network selects the best next action, while the target network provides a more stable estimate of its value, helping to mitigate optimistic bias during learning.

\vspace{0.2em}
\noindent\textit{Reward Model:}
The reward function incentivizes the agent to align scheduled resource allocations with link demand. The total reward is given by:
$$
R_{\text{scheduling}} = \sum_{i=1}^{N} \left( 1 - (w_i - \hat{w}_i)^2 \right)
$$
where $w_i$ denotes the normalized weight of the $i$-th link, and $\hat{w}_i$ is the weight assigned during scheduling.

The squared error penalizes large deviations, encouraging alignment with demand rather than just approximate coverage. The reward captures a global network objective by aggregating across all links, guiding the DDQN agent to learn scheduling policies that optimize overall utility while adhering to antenna and link activation constraints.

After the training converges, the agent generalizes to unseen topologies and traffic conditions. During inference, it ranks links by Q-value, applies the scheduling policy, and greedily activates links within per-node constraints, enabling scalable, low-latency scheduling in dynamic, interference-prone mmWave backhaul environments.

\subsection{DRL for Joint Bandwidth and Antenna Allocation}\label{sec:RL-band_antena}

To dynamically meet slice-level QoS requirements in congested mmWave IAB networks, we introduce a DRL-based agent that jointly allocates bandwidth and antenna resources. This allocation process occurs after link scheduling has been completed (as described in~\Cref{sec:RL-scheduling}) and operates on the residual resources—the bandwidth and antenna capacity remaining at BS1 and its neighbors.

The objective is to determine whether BS1, designated as persistently congested, should borrow wireless backhaul capacity and antenna resources from neighboring base stations~(BS2–BS7) to meet the demands of its active network slices over time intervals. These demands vary across slices and over time, and local resources may be insufficient due to sustained scheduling pressure. A dual-level DDQN agent jointly allocates bandwidth and antenna resources per slice among training episodes.

%Modeling both dimensions together allows the agent to account for their interdependence, ensuring, for example, that bandwidth is only borrowed when sufficient antenna capacity is also available. This coordination supports efficient, demand-aware offloading decisions.

\vspace{0.2em}
\noindent\textit{Observation model:}
At each decision point, the agent observes a vector:
$$
o_t = \left[ r_i^{\text{QoS}},\ b_0^{\text{avail}},\ b_j^{\text{avail}},\ a_0^{\text{avail}},\ a_j^{\text{avail}} \right], \quad j = 2,\dots,7
$$
Here, $r_i^{\text{QoS}} \in \mathbb{R}^2$ encodes the bandwidth and antenna requirements of slice $i$; $b_0^{\text{avail}}, a_0^{\text{avail}}$ represent BS1’s residual resources after scheduling; and $b_j^{\text{avail}}, a_j^{\text{avail}}$ are the remaining capacities at neighboring BSs. This compact representation enables the agent to assess local resource scarcity and offloading opportunities across the network.

\vspace{0.2em}
\noindent\textit{Action model:}
For each slice $i \in \{1,\dots,S\}$, the action space is defined by taking into account the traffic load on BS1’s wireless backhaul connection to the donor node, as well as the traffic load on the wireless links of the surrounding base stations. The action space is:
$$
\mathcal{A}_i = \{ e_1, e_2, \dots, e_7 \} \subset \mathbb{R}^7
$$
where each $e_j$ is a one-hot vector indicating the selected BS. Sequential allocation enables the agent to condition each decision on residual resource availability and prior slice allocations, which is critical under tight constraints.

\vspace{0.2em}
\noindent\textit{Reward model:}
To guide learning, the reward function promotes allocations that closely match each slice’s demand:
$$
R_{\text{resources}} = \sum_{i=1}^{S} \left[ 1 - (s_i - r_i)^2 \right]
$$
where $r_i$ is the demand for slice $i$, and $s_i$ is what the system borrowed from the selected neighbor. The reward is maximized when $s_i = r_i$ and penalized quadratically otherwise. This encourages the agent to minimize both under-provisioning, which degrades QoS, and over-provisioning, which wastes limited mmWave resources.

By allocating remaining resources in a fine-grained, context-aware manner, the agent enables slice-specific, congestion-responsive adaptation, enhancing service quality in dynamic IAB environments.

\section{Experimental Results and Analysis}\label{sec:evaluationresults}

\begin{table}[t]
\centering
\scriptsize
\begin{minipage}{\linewidth}
\centering
\captionof{table}{Training parameters for link scheduling and resource allocation.}
\begin{tabular}{p{0.6\linewidth} p{0.3\linewidth}}
    \toprule
    Learning rate ($\alpha$) (scheduling) & 0.001 \\
    Learning rate ($\alpha$) (resource) & 0.0001 \\
    Discount factor ($\gamma$) & 0.99 \\
    Initial exploration probability ($\varepsilon$) (scheduling) & 0.9 \\
    Initial exploration probability ($\varepsilon$) (resource) & 0.99 \\
    Minimum exploration probability ($\varepsilon_{\text{min}}$) & 0.01 \\
    Exploration decay rate (scheduling) & 0.995 \\
    Mini-batch size ($M$) (scheduling) & 32 samples \\
    Mini-batch size ($M$) (resource) & 64 samples \\
    Target network update interval ($C$) & Every 2 timesteps \\
    Experience replay buffer size ($N$) & 10{,}000 \\
    Optimization algorithm & Adam \\
    Number of training episodes ($E$) (scheduling) & 1000 \\
    Activation function (hidden layers) & ReLU \\
    eMBB slice requirements & (8000~MB, 20000~MB) \\
    URLLC slice requirements & (500~MB, 3000~MB) \\
    eMTC slice requirements & (500~MB, 2500~MB) \\
    Number of antennas required per slice & [1, 10] \\  
    Total BS Capacity  & 1~G \\ 
    \bottomrule
\end{tabular}
\label{tab:Training_parameters}
\vspace{1em}

\captionof{table}{Link scheduling performance.}
\begin{tabular}{p{0.42\linewidth} p{0.2\linewidth}}
    \toprule
    Predicted accuracy & 99.84\% \\
    Total inference time & 17 seconds \\
    Average inference time per graph & 0.17 seconds \\
    Achieved reward & 5457.09 \\
    \bottomrule
\end{tabular}
\label{tab:Link-scheduling_performance}
\vspace{1em}

\captionof{table}{Training performance summary for bandwidth and antenna allocation under different configurations.}
\begin{tabular}{lcc}
    \toprule
    \textbf{Parameter} & \textbf{Config 1} & \textbf{Config 2} \\
    \midrule
    Hidden Layers & 2 & 1 \\
    Episodes & 500 & 21 \\
    Exploration decay rate & 0.99 & 0.01 \\
    Bandwidth Convergence Episode & 357 & 7 \\
    Initial Bandwidth Reward & 261.98 & 262.56 \\
    Bandwidth Reward at Episode 100 & 275.31 & -- \\
    Maximum Bandwidth Reward & 281.19 & 281.19 \\
    Antenna Convergence Episode & 357 & 7 \\
    Initial Antenna Reward & 254.71 & 251.58 \\
    Antenna Reward at Episode 100 & 268.52 & -- \\
    Maximum Antenna Reward & 276.33 & 276.33 \\
    \bottomrule
\end{tabular}
\label{tab:training_summary}
\end{minipage}
\end{table}

\subsection{Network Simulation Setup}

To evaluate the proposed system, we simulate a dynamic mmWave IAB network over 24 hours, regenerating  network topologies randomly every 15 minutes to produce 96 distinct graph instances. The evaluation comprises three stages. First, the Greedy DDQN agent from \Cref{sec:RL-scheduling} is trained for link scheduling under hardware and traffic constraints. Second, the DRL framework from \Cref{sec:RL-band_antena} allocates bandwidth and antenna resources across three network slices. Third, system throughput and cumulative reward are analyzed under various configurations and compared to the ML benchmark in~\cite{morgado2023deep}. Experiments were run on a Dell desktop with a Core i9 processor, 12 vCPUs, 32GB RAM, and no GPU. Training parameters are detailed in Table~\ref{tab:Training_parameters}.

Each of the three network slices is assigned random, time-varying bandwidth and antenna requirements. With slice-level allocation occurring at each interval, this yields $96 \times 3 = 288$ selection events. To maintain consistency across episodes and metrics, all numerical features—link weights, throughput, and slice demands—are normalized to $[0, 1]$. Slice profiles are updated at every timestep to reflect realistic demand fluctuations. The maximum cumulative reward per slice (for bandwidth and antenna) is 288, representing perfect demand satisfaction at every step.

\vspace{0.2em}
\noindent\textit{Link Scheduling Parameters:}
The link scheduling agent uses a two-layer DDQN architecture with 32 ReLU-activated units per layer. More complex models with additional layers or neurons increased training time without notable performance gains, while a single-layer model underperformed in accuracy. The two-layer design was thus selected for its balance of simplicity and effectiveness.

\begin{figure}[t]
\centering
\includegraphics[width=.63\linewidth]{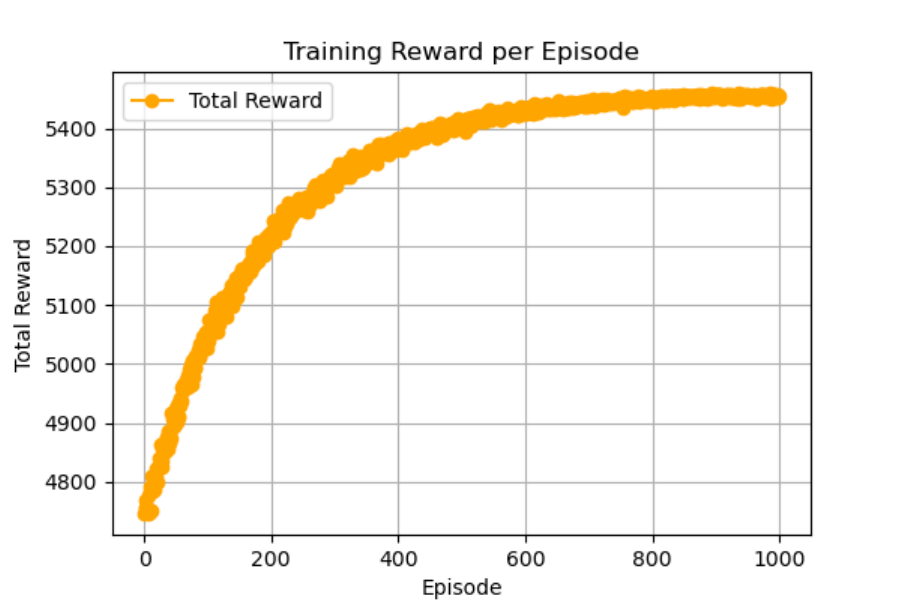} %.6
\caption{Total reward over 1000 episodes of link scheduling.}
\label{fig:link_scheduling_reward}
\end{figure}

%\begin{figure}[t]
%\centering
%\begin{minipage}{\linewidth}
%\centering

%\begin{subfigure}[t]{0.49\linewidth}
%    \centering
%    \includegraphics[width=\linewidth]{Figures/bandwidth-reward.pdf}
%    \caption{Bandwidth allocation}
%    \label{fig:Bandwidth_reward}
%\end{subfigure}
%\hfill
%\begin{subfigure}[t]{0.49\linewidth}
%    \centering
 %   \includegraphics[width=\linewidth]{Figures/antena_reward.pdf}
 %   \caption{Antenna allocation}
 %   \label{fig:Antena_reward}
%\end{subfigure}

%\caption{Achieved reward over 20 episodes for (a) bandwidth and (b) antenna allocation.}
%\label{fig:reward_comparison}
%\vspace{1em}

%\includegraphics[width=.55\linewidth]{Figures/Reward_link_scheduling.pdf}
%\caption{Total reward over 1000 episodes of link scheduling.}
%\label{fig:link_scheduling_reward}
%\end{minipage}
%\vspace{-0.25em}
%\end{figure}

\vspace{0.2em}
\noindent\textit{Resource Allocation Parameters:}
The bandwidth and antenna allocation agent uses a DDQN architecture trained under two configurations, summarized in Table~\ref{tab:training_summary}. Both use ReLU-activated hidden layers and a linear output layer for Q-value estimation. In the 500-episode setup, both agents were trained concurrently with epsilon gradually decaying to favor exploitation. To analyze exploration dynamics, we also evaluate a short-horizon setup with a 0.01 decay rate over 21 episodes, similar to the benchmark~\cite{morgado2023deep}. Despite limited exploration, the agent quickly converges to near-optimal performance.

\subsection{Performance Assessment}

\subsubsection{Link Scheduling Performance}

Figure~\ref{fig:link_scheduling_reward} shows the total reward over 1000 episodes, with our system reaching near-optimal performance by episode 902. It generalizes well across 96 unseen topologies with varied interference and node associations, achieving 99.84\% scheduling accuracy. With an average inference time of 0.17 seconds per graph, the system demonstrates high scheduling efficiency and real-time suitability (Table~\ref{tab:Link-scheduling_performance}). Scheduled link activations enable precise residual bandwidth and antenna resource tracking for the next allocation stage.

\subsubsection{Bandwidth and Antenna Allocation Performance}

Training performance under two resource allocation configurations is summarized in Table~\ref{tab:training_summary}. Our design achieves 97.64\% of the maximum possible reward, with a total of 281.19 out of 288, indicating near-optimal performance. Configuration 1, with two hidden layers, was chosen for its consistent stability across runs and required 46,279.62 seconds over 500 epochs. Configuration 2 achieved optimal reward faster, completing training in 2,154.97 seconds over 21 epochs. Although both configurations performed equivalently, Configuration 1 was more stable, while Configuration 2 offered a faster alternative when training time is critical. These results highlight the robustness of the allocation policy under realistic load conditions and variable slice demands.

%Reward trends for each configuration are visualized in Figure~\ref{fig:reward_comparison}, confirming fast convergence and consistent allocation performance across bandwidth and antenna dimensions.

\subsubsection{Throughput and Reward Analysis Against Benchmark}

\begin{figure}[t]
\centering
\begin{minipage}{\linewidth}
\centering
\includegraphics[width=.67\linewidth]{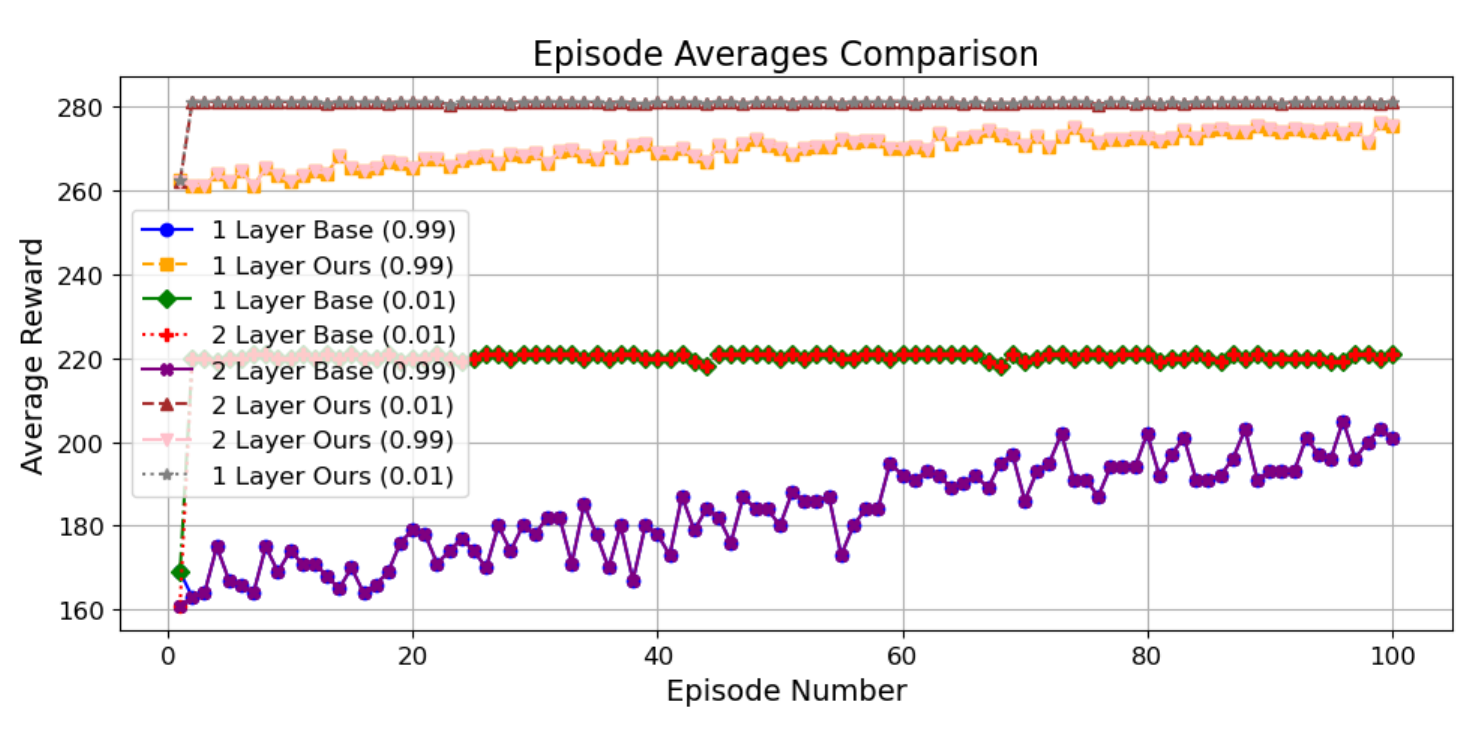} %0.76
\captionof{figure}{Reward comparison across configurations.}
\label{fig:Reward_all}
\vspace{1.5em}

\includegraphics[width=.65\linewidth]{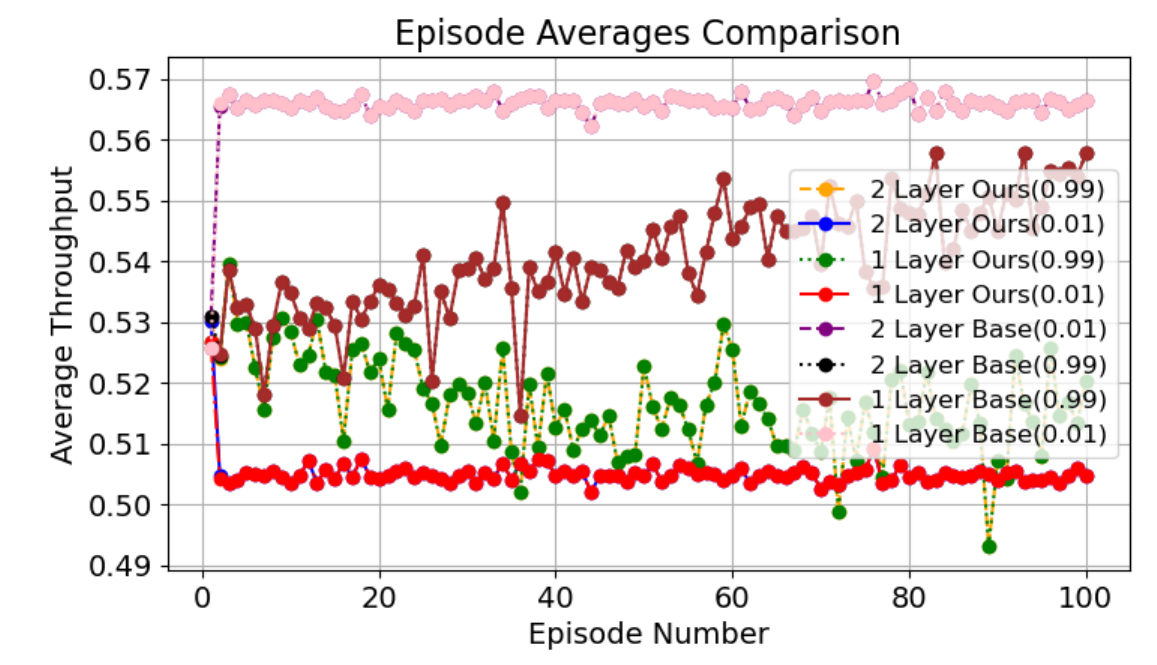} %0.56
\captionof{figure}{Throughput comparison across configurations.}
\label{fig:fig_throuphput}
\end{minipage}
\vspace{-0.25em}
\end{figure}

We benchmark our approach against the ML-based method from \cite{morgado2023deep}, which uses learned policies for resource assignment without accounting for residual capacity or link quality. Reward comparisons across configurations are shown in Figure~\ref{fig:Reward_all}, and resulting throughput in Figure~\ref{fig:fig_throuphput}.

Our system outperforms the ML benchmark from \cite{morgado2023deep} in both cumulative reward and throughput allocation. To better reflect real-world conditions, we modify the benchmark’s reward mechanism to grant a reward of +1 when allocated bandwidth meets or exceeds a slice's demand—unlike the original design, which rewards only exact matches. This adjustment allows for fairer evaluation under dynamic network scenarios. In a representative case, with slice demands ${0.81, 0.54, 0.22}$ and available bandwidths at seven base stations (BSs) as ${0.50, 0.39, 0.65, 0.75, 0.41, 0.37, 0.52}$, our system selects BS4, BS7, and BS6, achieving a cumulative reward of approximately 2.97. The benchmark, constrained by less adaptive criteria, selects BS3 and BS4 and obtains a slightly lower reward just above 2. In this scenario, our system allocates 1.64 units of bandwidth to satisfy 1.57 units of demand, wasting only 0.07 units. In contrast, the benchmark allocates 1.4 units inefficiently. Across multiple tests, our system consistently outperforms the benchmark, demonstrating more efficient bandwidth utilization and strong potential for real-world deployment.

As shown in Figure~\ref{fig:Reward_all}, our system, using both one- and two-layer configurations with a decay rate of 0.01, converges faster and achieves a higher cumulative reward of 281.19 (gray, brown), compared to the benchmark’s 221 (green). This corresponds to 97.64\% of the maximum possible reward, versus 76.74\% for the benchmark—a 20.90\% improvement driven by greater adaptability to dynamic conditions. Similarly, Figure~\ref{fig:fig_throuphput} shows that our one-layer model with a 0.01 decay rate (red) consistently outperforms the benchmark (pink), despite both using similar architectures. Lower throughput values indicate better alignment with slice demands and more efficient resource use. While deeper models or slower decay rates (e.g., 0.99) slow convergence, the final allocation quality remains consistent, confirming the robustness of our approach across training conditions.

\subsubsection{Results Summary} 
In summary, our framework achieves up to 99.84\% link scheduling accuracy, outperforms the benchmark by up to 20.90\% during bandwidth and antenna allocation, and enables scalable, slice-aware resource optimization in dense, dynamic IAB deployments. These results highlight the effectiveness of integrating DRL-based scheduling with resource-aware slice provisioning. By adapting to traffic variability and hardware constraints, the framework offers a practical solution for efficient, autonomous operation in real-world IAB networks.

\section{Conclusion and Future Work}\label{sec:conclusion}

We proposed a DRL-based framework for joint link scheduling and resource slicing in mmWave-enabled IAB networks. By combining a greedy Double Deep Q-Network (DDQN) scheduler with a multi-agent DDQN bandwidth and antenna allocator, the system enables adaptive, fine-grained control under strict bandwidth and antenna constraints. Preliminary evaluations on dynamic topologies show near-optimal scheduling accuracy and significant throughput gains over baseline methods. Future work will examine techniques for generalization to larger-scale networks, non-LoS conditions, and integration with lightweight security mechanisms.

\bibliography{references}

\begin{thebibliography}{10}
\providecommand{\url}[1]{#1}
\csname url@samestyle\endcsname
\providecommand{\newblock}{\relax}
\providecommand{\bibinfo}[2]{#2}
\providecommand{\BIBentrySTDinterwordspacing}{\spaceskip=0pt\relax}
\providecommand{\BIBentryALTinterwordstretchfactor}{4}
\providecommand{\BIBentryALTinterwordspacing}{\spaceskip=\fontdimen2\font plus
\BIBentryALTinterwordstretchfactor\fontdimen3\font minus
  \fontdimen4\font\relax}
\providecommand{\BIBforeignlanguage}[2]{{%
\expandafter\ifx\csname l@#1\endcsname\relax
\typeout{** WARNING: IEEEtran.bst: No hyphenation pattern has been}%
\typeout{** loaded for the language `#1'. Using the pattern for}%
\typeout{** the default language instead.}%
\else
\language=\csname l@#1\endcsname
\fi
#2}}
\providecommand{\BIBdecl}{\relax}
\BIBdecl

\bibitem{sadovaya2024impact}
Y.~Sadovaya, D.~Moltchanov, W.~Mao, S.-p. Yeh, O.~Semiari, H.~Nikopour,
  S.~Talwar, and S.~Andreev, ``{Impact of System-Specific Factors on Scheduling
  and Resource Allocation in mmWave IAB Networks},'' in \emph{{ICC 2024-IEEE
  International Conference on Communications}}.\hskip 1em plus 0.5em minus
  0.4em\relax IEEE, 2024.

\bibitem{ghodhbane2024delay}
C.~Ghodhbane, P.~Savelli, C.~Gueguen, and X.~Lagrange, ``{Delay reduction and
  interference mitigation by resource management for IAB networks},''
  \emph{{Discover Applied Sciences}}, vol.~6, no.~12, p. 625, 2024.

\bibitem{song2017relay}
S.~Song, Y.~Liu, Q.~Song, and L.~Guo, ``{Relay selection and link scheduling in
  cooperative free-space optical backhauling of 5G small cells},'' in
  \emph{{2017 IEEE/CIC International Conference on Communications in China
  (ICCC)}}.\hskip 1em plus 0.5em minus 0.4em\relax IEEE, 2017.

\bibitem{cho2025resource}
C.-W. Cho and M.-S. Pan, ``{Resource Scheduling in MU-MIMO and NOMA Enabled
  Integrated Access and Backhaul Networks},'' \emph{{IEEE Open Journal of the
  Communications Society}}, 2025.

\bibitem{ganesh2024impact}
D.~Ganesh, R.~Upadhyay, and A.~K. Saxena, ``{Impact of Antenna Beam Patterns on
  the Performance of Satellite Communication Links},'' in \emph{{2024
  International Conference on Optimization Computing and Wireless Communication
  (ICOCWC)}}.\hskip 1em plus 0.5em minus 0.4em\relax IEEE, 2024.

\bibitem{morgado2023deep}
A.~J. Morgado, F.~B. Saghezchi, P.~Fondo-Ferreiro, F.~Gil-Casti{\~n}eira,
  M.~Papaioannou, K.~Ramantas, and J.~Rodriguez, ``{Deep Reinforcement Learning
  for Backhaul Link Selection for Network Slices in IAB Networks},'' in
  \emph{{GLOBECOM 2023-2023 IEEE Global Communications Conference}}.\hskip 1em
  plus 0.5em minus 0.4em\relax IEEE, 2023.

\bibitem{shashidhara2024hierarchical}
K.~Shashidhara, R.~Srivastava, R.~A. Sharief, and N.~Athmika, ``{Hierarchical
  Scheduling of Flexible 5G Networks},'' in \emph{{2024 First International
  Conference on Software, Systems and Information Technology (SSITCON)}}.\hskip
  1em plus 0.5em minus 0.4em\relax IEEE, 2024.

\bibitem{vu2016joint}
T.~K. Vu, M.~Bennis, S.~Samarakoon, M.~Debbah, and M.~Latva-aho, ``{Joint
  in-band backhauling and interference mitigation in 5G heterogeneous
  networks},'' in \emph{{European Wireless 2016; 22th European Wireless
  Conference}}.\hskip 1em plus 0.5em minus 0.4em\relax VDE, 2016.

\bibitem{qian2019joint}
Z.~Qian, Y.~Yang, K.~Srinivasan, and N.~B. Shroff, ``{Joint antenna allocation
  and link scheduling in FlexRadio networks},'' in \emph{{IEEE INFOCOM
  2019-IEEE Conference on Computer Communications}}.\hskip 1em plus 0.5em minus
  0.4em\relax IEEE, 2019.

\bibitem{morgado2024intelligent}
A.~J. Morgado, F.~B. Saghezchi, P.~Fondo-Ferreiro, F.~Gil-Casti{\~n}eira, and
  J.~Rodriguez, ``{Intelligent Backhaul Link Selection for Traffic Offloading
  in B5G Networks},'' \emph{IEEE Access}, 2024.

\bibitem{fabian2021performance}
P.~Fabian, G.~Z. Papadopoulos, P.~Savelli, and B.~Cousin, ``{Performance
  evaluation of integrated access and backhaul in 5G networks},'' in
  \emph{{2021 IEEE Conference on Standards for Communications and Networking
  (CSCN)}}.\hskip 1em plus 0.5em minus 0.4em\relax IEEE, 2021.

\end{thebibliography}
\bibliographystyle{IEEEtran}

\end{document}